\documentclass[format=acmsmall, review=false, screen=true]{acmart}

\usepackage{amsmath,amssymb,amsfonts}
\usepackage{algorithmic}
\usepackage{graphicx}
\usepackage{textcomp}
\usepackage{multicol}

\usepackage{mathtools}
\usepackage{amsmath,amssymb}
\usepackage{amsfonts}
\usepackage{braket}

\makeatletter
\let\@authorsaddresses\@empty
\makeatother

\makeatletter
\renewcommand\@formatdoi[1]{\ignorespaces}
\makeatother
\setcopyright{none}
\renewcommand\footnotetextcopyrightpermission[1]{}
\begin{document}

\title{Solving Multi-Coloring Combinatorial Optimization Problems Using Hybrid Quantum Algorithms}

\author{Young-Hyun Oh}
\email{ohy@us.ibm.com}
\affiliation{%
  \institution{IBM}
  \streetaddress{4205 S Miami Blvd.}
  \city{Durham}
  \state{North Carolina}
  \postcode{27703}
}
\author{Hamed Mohammadbagherpoor}
\email{hmohamm2@ncsu.edu}
\affiliation{%
  \institution{North Carolina State University}
  \streetaddress{890 Oval Dr.}
  \city{Raleigh}
  \state{North Carolina}
  \postcode{27606}
}
\author{Patrick Dreher}
\email{padreher@ncsu.edu}
\affiliation{%
  \institution{North Carolina State University}
  \streetaddress{890 Oval Dr.}
  \city{Raleigh}
  \state{North Carolina}
  \postcode{27606}
}
\author{Anand Singh}
\email{anand.singh@ibm.com}
\affiliation{%
  \institution{IBM}
  \streetaddress{4205 S Miami Blvd.}
  \city{Durham}
  \state{North Carolina}
  \postcode{27703}
}
\author{Xianqing Yu}
\email{xianqing@us.ibm.com}
\affiliation{%
  \institution{IBM}
  \streetaddress{4205 S Miami Blvd.}
  \city{Durham}
  \state{North Carolina}
  \postcode{27703}
}
\author{Andy J. Rindos}
\email{rindos@us.ibm.com}
\affiliation{%
  \institution{IBM}
  \streetaddress{4205 S Miami Blvd.}
  \city{Durham}
  \state{North Carolina}
  \postcode{27703}
}

\renewcommand{\shortauthors}{Young and Hamed, et al.}

\begin{abstract}
The design of a good algorithm to solve NP-hard combinatorial approximation problems requires specific problem domain knowledge and often needs a trial-and-error problem solving approach. Graph coloring is one approach that can provide an efficient solution for certain types of combinatorial applications. Some optimization algorithms have been proposed to solve these type of multi-coloring graph problems such as flight scheduling, frequency allocation in networking, and register allocation.  For many of these examples a simple searching method to find an optimal solution can be the best option. Although this approach may provide a path to a solution for small sized problems, the computation cost will increase exponentially as the graph size and the number of colors increase. To mitigate such intolerable overhead, we present a hybrid quantum computing approach for multi-coloring graph problems that utilizes the variational quantum eigensolver (VQE) and quantum approximate optimization algorithm (QAOA) techniques. Our approach transforms each combinatorial problem into a corresponding Ising model formulation by using the calculated Hamiltonian matrices that are fed into the VQE and QAOA algorithms to produce the best solutions in polynomial time. We have applied this approach to three combinatorial applications and the implementation results demonstrate that our hybrid approach can find an optimal solution for each problem. The proposed technique can be applicable for finding optimal solutions to other combinatorial problems in various fields. 
\end{abstract}


\maketitle
\thispagestyle{empty}
\pagestyle{plain}

\section{Introduction}
Quantum computers appear to be useful for solving some types of problems that can take exponential time to converge into the correct solution~\cite{ref1}. Shor's algorithm~\cite{ref1}, Deutsch-Jozsa algorithm~\cite{ref43}, Grover's algorithm~\cite{ref2} and quantum phase estimation (QPE)~\cite{ref44, ref45} are the well-known quantum algorithms that can demonstrate the advantages of quantum computers for solving several classical problems efficiently compared to classical computers by utilizing quantum properties of superposition and entanglement.  
The first generation of noisy intermediate-scale quantum (NISQ)~\cite{ref3} computers can now provide a framework to re-formulate algorithms originally optimized for digital computers into a new form that can be suitable for the quantum computing hardware platforms. These re-formulations may help the problem to run exponentially faster on a quantum computer. These framework reformulation may also help provide insights and be applied to explore sophisticated fields that are currently inaccessible to find a solution from the most powerful high-performance machines~\cite{ref4}. 

A key difference between an algorithm formulated for digital computers and an algorithm designed for quantum computers is that digital computations are modeled on a Load-Run-Read cycle while quantum computers operate on a Prepare-Evolve-Measure cycle. The information flow for digital algorithms assumes that digital bit input data is inserted into the system, the program executes, and finally the output is read. In a quantum computer the qubit states are prepared as input, manipulation of the input states is done using quantum operators, and then the results are measured~\cite{ref5}. 

Two important capabilities that are available to the quantum computer developer to strive for a quantum advantage over their digital counterparts are the quantum mechanical properties of both of superposition and entanglement of the qubits~\cite{ref6}. For example, quantum computers can be capable to simultaneously evaluate all the combinatorial states by taking the advantages of quantum computing properties but classical computers need to evaluate the states individually. This type of evaluation can be applied to most of the artificial intelligent (AI) problems in which the algorithm should be tested and analyzed during the evaluation process.

In recent years remarkable progress has been made developing quantum computing hardware platforms and several companies have built platforms that researchers can use to run their quantum algorithms on NISQ devices (ex. IBM Q machines~\cite{ref32}, Rigetti QPU~\cite{ref33}, trapped ion quantum computers~\cite{ref34}, and D-Wave quantum annealing machines~\cite{ref35}). However, running quantum algorithms on these machines is challenging because the quantum computers cannot produce sufficiently reliable results due to high error rates, limited circuit depth, a low number of reliable qubits, and the limitation of applicable quantum algorithms to solve their complicated problems. 
These companies have also developed quantum simulators to assist researchers in re-formulating problems in ways that run on their quantum computing hardware platforms. Although the quantum simulator can mimic quantum computing hardware behavior, simulator utilization requires that researchers divide the problem into smaller problem sets because the simulator must run on a classical computer. The ability to use these simulators is dependent on host computer's physical resources, limiting the problem size that can be implemented.  

Capitalizing on the access to both quantum simulators and NISQ hardware platforms multiple near-term algorithms have been developed to run hybrid quantum-classical approaches. Variational quantum eigensolver (VQE) and quantum approximate optimization algorithm (QAOA)~\cite{ref7,ref8} represent  state-of-the-art algorithms for using NISQ devices and classical computing resources. In these algorithms, a quantum computer is initialized at an arbitrary state and then allowed to evolve in time. Classical optimizers are applied at each step to update the quantum circuit parameters. These hybrid approaches require no error correction techniques in order to run quantum circuits on the NISQ devices. The quantum circuits can be constructed by parameterizing the gates such as rotation around $Z$ axis. A similar VQE/QAQA approach has been used for transactional settlements in financial applications using a quantum mixed binary optimization approach~\cite{ref42}. 

In this paper, we aim to comprehensively investigate the design, development, and application of VQE and QAOA algorithms to graph theory for solving a $k$-coloring graph problem. We present a method in which the $k$-coloring problem is mapped into a combinatorial optimization problem. The problem is converted to the appropriate Hamiltonian matrix that can then be used by quantum computers to find an optimum solution. Different quantum circuits are considered for testing and analyzing the proposed technique.  In order to converge into the correct solution, classical optimizers are applied to the measured quantum computers results. This proposed technique is tested by applying it to three different $k$-coloring graph problems to find an optimal solution for each.  

The paper is organized as follows: Section~\ref{section-2} describes the quantum variational eigensolver (VQE) technique and follows the explanation of the classical optimization process. In section~\ref{section-3}, a combinatorial optimization approach is presented and the converting method from a classical problem to a Hamiltonian matrix is explained. Section~\ref{section-4} describes the classical methods to solve the $k$-coloring problems. Section~\ref{section-5} represents the hybrid classical and quantum approaches that can be used to solve multi-coloring graph problems. In section~\ref{section-6}, the QAOA algorithm is applied to different combinatorial problems and its performance is evaluated. Finally, section~\ref{section-7} summarizes the results and conclusions from this work.

\section{Variational quantum eigensolver (VQE) }\label{section-2}

Quantum computers can be suitable for solving specific problems exponentially faster than classical computers. However, achieving this goal for some large-scale problems can take several decades and may require breakthrough technologies due to the high error rates present in today's NISQ quantum computers. Because of these hardware limitations, most quantum algorithms are not be able to find correct or optimal solutions. To mitigate these limitations, researchers have been investigating hybrid algorithms that can be run efficiently using a combination of both quantum and classical computers. Eigenvalues, eigenvectors, and optimization problems play important roles in such hybrid algorithms. 

Peruzzo and McClean~\cite{ref7} presented the variational quantum eigensolver (VQE) algorithm to solve the eigenvalues and binary optimization problems that cannot be solvable by solely using classical computers. In this technique, both classical and quantum computers are being used to converge intermediate results in the correct solution. The quantum state corresponding to the minimum energy of the Hamiltonian matrix can be solved by applying a quantum/classical hybrid variational technique. This algorithm was tested both on a photonic quantum chip and ion trap quantum computers. 

In the VQE approach~\cite{ref7}, the Hamiltonian matrix $H$ can be derived by the interaction spins or electronic systems in physical systems. The matrix $H$ can be calculated to find the eigenvectors $\ket{\psi_i(t)}$ and the corresponding eigenvalues $\lambda_i$ of the Hamiltonian matrix by converting an optimization problem into the adiabatic quantum computation. The main goal of VQE algorithm is to find the desired eigenvectors and eigenvalues by optimizing the designed parameters inside the quantum circuits. The lowest eigenvalue is the primary interest of the VQE approach since it is the optimum solution of the problem, the same as the lowest energy is the solution to the Hamiltonian matrix in physical systems. The VQE technique can be used to find the solution for Eq. (\ref{VQE}),
\begin{equation}\label{VQE}
\underset{\theta} min\bra{\psi(\theta)}H\ket{\psi(\theta)}.
\end{equation}
Because $\ket{\psi(\theta)}$ is the normalized solution for Eq. (\ref{VQE}), it can be the minimum eigenvalue of Hamiltonian matrix $H$. The $H$ matrix is constructed based on the summation of the tensor products of Pauli operators and $H$ is $H = \sum_{i} \beta_i H_i$ where $H_i$ are the tensor products of Pauli operators and $\beta_i$ are the real coefficients. 
By considering this reformulation for $H$ matrix, Eq.~(\ref{VQE}) can be represented as,
\begin{equation}\label{VQE2}
E(\theta) = \underset{\theta} min\bra{\psi(\theta)}H\ket{\psi(\theta)} = \underset{\theta} min \sum_{i} \beta_i \bra{\psi(\theta)}H_i\ket{\psi(\theta)}.
\end{equation}

\begin{figure}[!t]
  \begin{center}
  \includegraphics[width=3.5in]{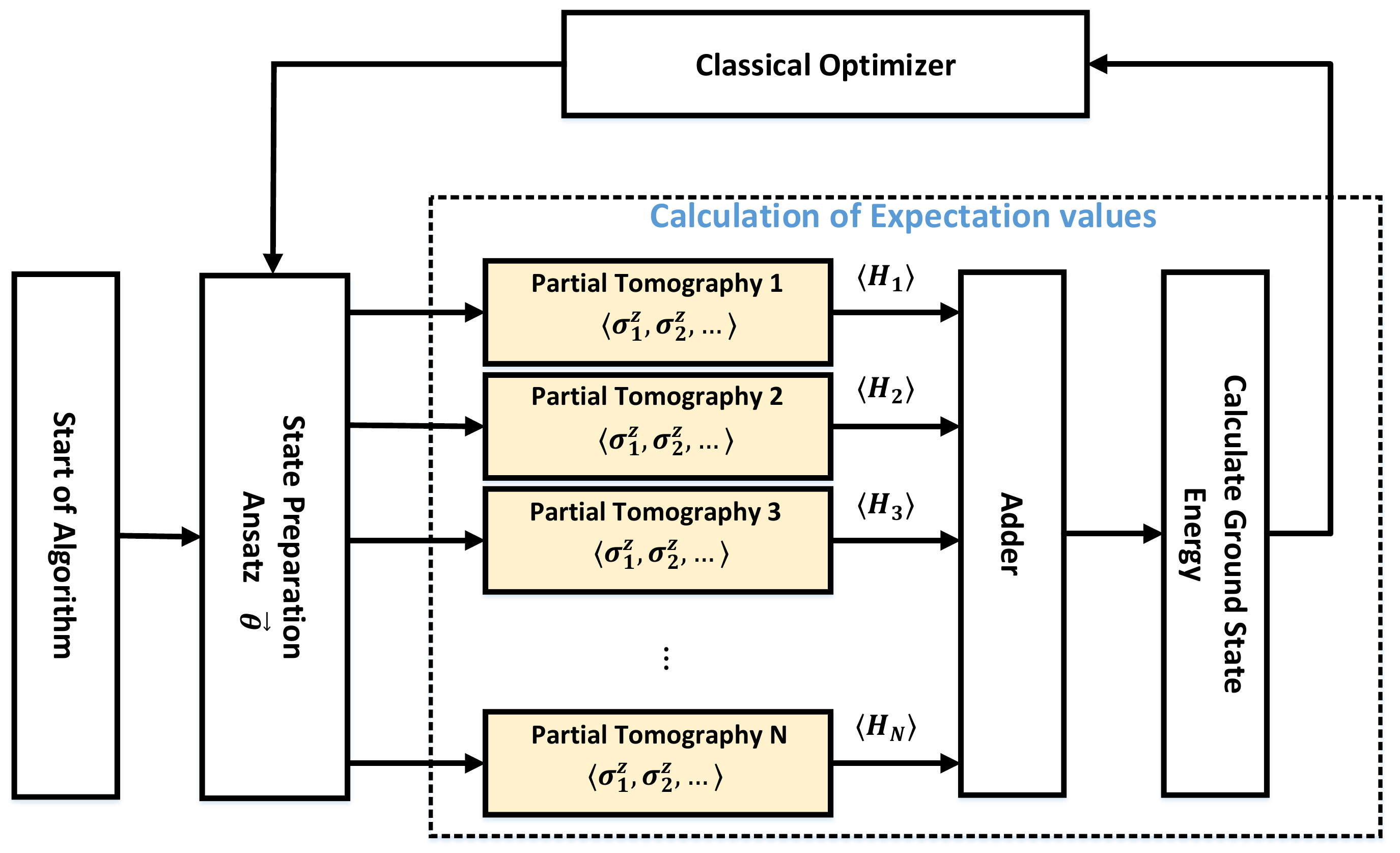}\\
  \caption{The schematic diagram of the variational quantum eigensolver (VQE) algorithm
}
  \label{VQE_generalview}
  \end{center}
\vspace{-1em}
\end{figure}
Finding the minimum eigenvalue can be an NP-hard problem as the size of the Hamiltonian matrix exponentially increases. 
To address this problem, different quantum techniques have been developed to find the minimum eigenvalue of the Hamiltonian matrix~\cite{ref22}. Quantum phase estimation (QPE) is one of the main techniques in which it can be applied to the unitary quantum circuit derived from a problem to find the minimum eigenvalue of the unitary operator that is the solution of the Eq.~(\ref{VQE}). However, implementing QPE algorithm requires a large number of controlled rotation gates so that it is not trivial or possible to find the correct answer to a problem using near-term noisy quantum computers. 

Alternatively, researchers have been developing some hybrid classical-quantum algorithms (e.g., VQE and QAOA) to solve a problem by finding the minimum eigenvalue of the Hamiltonian Matrix in a recursive optimization process using both NISQ quantum computers and classical computing resources. The optimization of a Hamiltonian matrix $H$ can be determined by applying a hybrid classical-quantum technique. In the hybrid system, a classical computer first finds the optimum values for variational parameters, $\theta$. Second, a quantum computer uses the $\theta$ to calculate the expectation value ($\bra{\psi(\theta)}H_i\ket{\psi(\theta)}$). Then, the calculated expectation value will be fed back into a classical algorithm to update the parameter $\theta$. These steps will be recursively processed until the hybrid system finds the minimum eigenvalue of the $H$ matrix~\cite{ref23}. 

In this hybrid system approach, all the qubits in the quantum computer are initialized at zero states $\ket{0}$ and then a parameterized quantum circuit ($U(\theta)$) known as an ansatz will be applied to the qubits. As a result the expectation values will be calculated and used in the classical optimization process~\cite{ref8}. Choosing an ansatz operator as a variational form plays an important role in the optimization process. In most cases, an ansatz should be able to efficiently run in quantum devices. Different classical optimizers can be used to solve the optimum values for $\theta$. These techniques perform various forms of gradient descent methods to apply iterative optimization algorithms to solve the problem~\cite{ref16,ref17}. 

This paper considers three well-known classical optimizers to optimize variational parameters. First, the constrained optimization by linear approximation (COBYLA) algorithm is a numerical optimization method that finds the min/max of the objective function without knowing the gradient value~\cite{ref41}. Second, the limited memory Broyden - Fletcher - Goldfarb - Shanno (L-BFGS) algorithm is a quasi-Newton optimization method that is mainly used for the parameter identification in machine learning techniques with limited amounts of computer memory~\cite{ref40}. Last, the sequential least squares programming (SLSQP) algorithm is an iterative optimization algorithm that is used for constraint nonlinear problems and uses the Han–Powell quasi–Newton method to find the min/max of the objective function~\cite{ref39}.    

Fig.~\ref{VQE_generalview} describes the general view of the VQE algorithm. As indicated in the figure, all of the qubits in the quantum computer are initialized to the zero state (i.e., state preparation).  The variational quantum circuit is then applied to the qubits after the initialization step and then outputs are measured.  Next, the results are then fed back into the classical computer to update the variational parameters and this process is repeated until the VQE algorithm finds the minimum eigenvalue of the problem.

\section{Binary optimization problem and quantum Hamiltonian}\label{section-3}
Combinatorial optimization (CO) is one of the essential optimization problems that can be used and applied in most of the industry and research applications. In particular, the CO for the graph coloring problems has been extensively used in different application domains such as transportation management, social media, telecommunication, and resource scheduling. Most of these types of problems are NP-hard problems that can be attractive but challenging for researchers to solve. 

Three different types of approaches can be mainly considered to tackle the CO problems: the exact solution algorithm, approximation algorithm, and heuristic algorithm~\cite{ref36}. The exact algorithm uses the enumeration with integer formulation to find the exact solution but it might not be able/possible to find the exact answer as the size of the problem increases. On the other hand, the approximation algorithm might find a solution to the problem in polynomial time. However, there is no guarantee that the results of the algorithm converge into a minimum or optimal solution.  There are situation where the results may display a local minimum that requires an an exponential increase in computational overhead as the problem size increases. The heuristic algorithm can be used for some specific types of CO problems. These algorithms are simple and fast because they use the trial and error approach to find a solution but they are lack of theoretical guarantee to find a correct answer. Moreover, in some cases, finding a good solution for this type of problem can be almost infeasible. 

Recently, Quadratic Unconstrained Binary Optimization (QUBO) has been introduced to cover a large variety of CO problems~\cite{ref21}. The QUBO technique can find an optimal solution for many types of NP-hard problems such as graph partitioning, graph coloring, scheduling management, and register allocation. 
A QUBO problem can be formulated by Eq. (\ref{QUBO}) as described in~\cite{ref31},
\begin{equation}\label{QUBO}
 J = min \;\;\{X^T Q X + g^T X  + c\},
\end{equation}

\noindent where $Q$ is a $n\times n $ matrix with real numbers, and $X$ is the state vector with element of $\{0,1\}^n$. The goal of Eq. (\ref{QUBO}) is to find the minimum of the objective function which is the minimum eigenvalue of the matrix $H$. Eq. (\ref{QUBO}) can also be expressed by Eq. (\ref{QUBO_2}) as described in \cite{ref19}, 
\begin{equation}\label{QUBO_2}
 \sum_{i,j} W_{ij} x_i (1-x_j) + \sum_{i} W_i x_i,
\end{equation}

\noindent where $W_i$ are the linear weights $Q_{ii}$ for $i=1,2,...,n$ and $W_{ij}$ are the quadratic weights $Q_{ij}$ for $i<j$. Here $x_i$ is a binary variable that can take $0$ or $1$ and as a result $x_i^2 = x_i$. In order to transfer Eq. (\ref{QUBO_2}) into an Ising model, the following substitution can be used as 
\begin{equation}\label{QUBO_sub}
 x_i = \frac{Z_i + 1}{2},
\end{equation}

\noindent where $ x_i\in\{0,1\}$ and $Z_i \in \{-1,1\}$ for $i=0,...,n$.
By applying the substitution the Hamiltonian matrix would be \cite{ref18, ref20},
\begin{equation}\label{QUBO_sub}
\begin{multlined}
 \sum_{i,j} \frac{W_{ij}}{4} (1-Z_i) (1+Z_j) + \sum_{i} \frac{W_i}{2} (1-Z_i) \\=
 -\frac{1}{2} \sum_{i,j} W_{ij} Z_i Z_j + \sum_{i} W_{i} Z_i + c.
 \end{multlined}
\end{equation}
This Hamiltonian matrix will be a form of tensor products of Pauli $Z$ operators. Note that since the $Z$ is a diagonal matrix, the Hamiltonian matrix should be diagonal so that it can be used in VQE optimization process.

\section{Graph theory and $k$-coloring problem }\label{section-4}

The main goal of graph coloring problems is to find the optimal number of colors and to reduce the coloring time. Graph coloring techniques can be vastly used in many practical applications such as scheduling and timelabling, biprocessor tasks, optimizing combustible chemical combinations, register allocation, routing assignments, and frequency assignment to the channels~\cite{ref12,ref13}. 

In graph theory, a graph $G(V,E)$ consists of a vertex set $V(G) = \{1,2,...,n\}$ and edges $E(G) = \{e_{ij}, e_{i,j}\in V(G)\}$ where $e_{ij}$ is the link between the vertex $i$ and $j$. Graphs with the nodes connected to the other nodes based on the problem constraints are defined as connected graphs.  These graphs can be directed where the link between the nodes shows the direction of the edge. 

In this paper, we consider a $k$-coloring problem as an undirected graph. A $k$-coloring problem is an approach that assigns $k$ colors to the vertices such that the colors for the adjacent vertices are not the same as each other. Many practical problems can be categorized in these graph coloring formulations such as flight gate allocation, frequency assignment for the substation channels, and register allocation.

Multi-coloring graph problems assign $k$ colors to all the nodes in a graph in which connected nodes (i.e., adjacent nodes) should not have the same color each other. The graph coloring problem is categorized as an NP-hard problems and the $k$-coloring graph problem is NP-complete for any integer $k\geq3$~\cite{ref9, ref10}. 

The chromatic number $\chi(G)$ of the graph ($G$) is defined as the smallest number of colors that can be used for coloring the graph. Determining the chromatic number is also an NP-hard problem and even approximating the number itself is a difficult problem in graph theory. Different heuristics algorithms such as the first fit (FF), saturation degree order (SDO) and largest degree order (LDO) have been proposed to satisfy the $k$-coloring constraints by selecting appropriate colors for the vertices in a graph~\cite{ref10,ref11}. A combination of the heuristic algorithm is used to reduce the processing time and to minimize the number of assigned colors. Due to the enormous computation overhead finding the chromatic number for a large graph is not possible. The known upper bound for finding a chromatic number can be expressed by $\chi(G) \leq \sqrt{2m}+1$ where $m$ is the number of edges in the graph. 

Alternatively, a greedy heuristic algorithm is a simple and classical approach to find a sub-optimal solution for proper coloring of the graph. The algorithm tries to solve the problem step by step by assigning a color to the given node~\cite{ref14,ref15}. However, this algorithm requires more number of colors than the chromatic number of the graph. Local search is still a powerful technique to solve the multi-coloring problems.

Although there are some algorithms proposed to solve multi-coloring problems, the simple searching may still be the best approach for finding an optimal solution. However, this approach can become exponentially difficult as the size of the graph and the number of colors increase. Using quantum computers may provide an advantage for finding a good solution for these types of graph coloring problems in polynomial time.  Succeeding with such an approach will be considerably beneficial in the field of quantum optimization research.

\subsection{Classical algorithms for graph coloring problems}
This section presents two different classical techniques: a greedy coloring algorithm and a backtracking $k$-coloring algorithm for coloring a graph.
\subsubsection{Greedy algorithm}
A greedy algorithm assigns positive numbers to the colors and uses a new numbered-color for a node each time~\cite{ref37}. The algorithm requires three steps to be applied to vertices in a graph. First, a vertex should be colored with the first/lowest number of colors (e.g., color \#1).  Second, another uncolored vertex should select the second-lowest number of colors to be assigned to the vertex. (Note that the selected color should not be used in the previous step for the adjacent vertices.) Third,  the step 1) and 2) will be continuously repeated until all vertices are colored. 

\begin{figure}[t]
  \begin{center}
  \includegraphics[width=2.5in,height=1.8in]{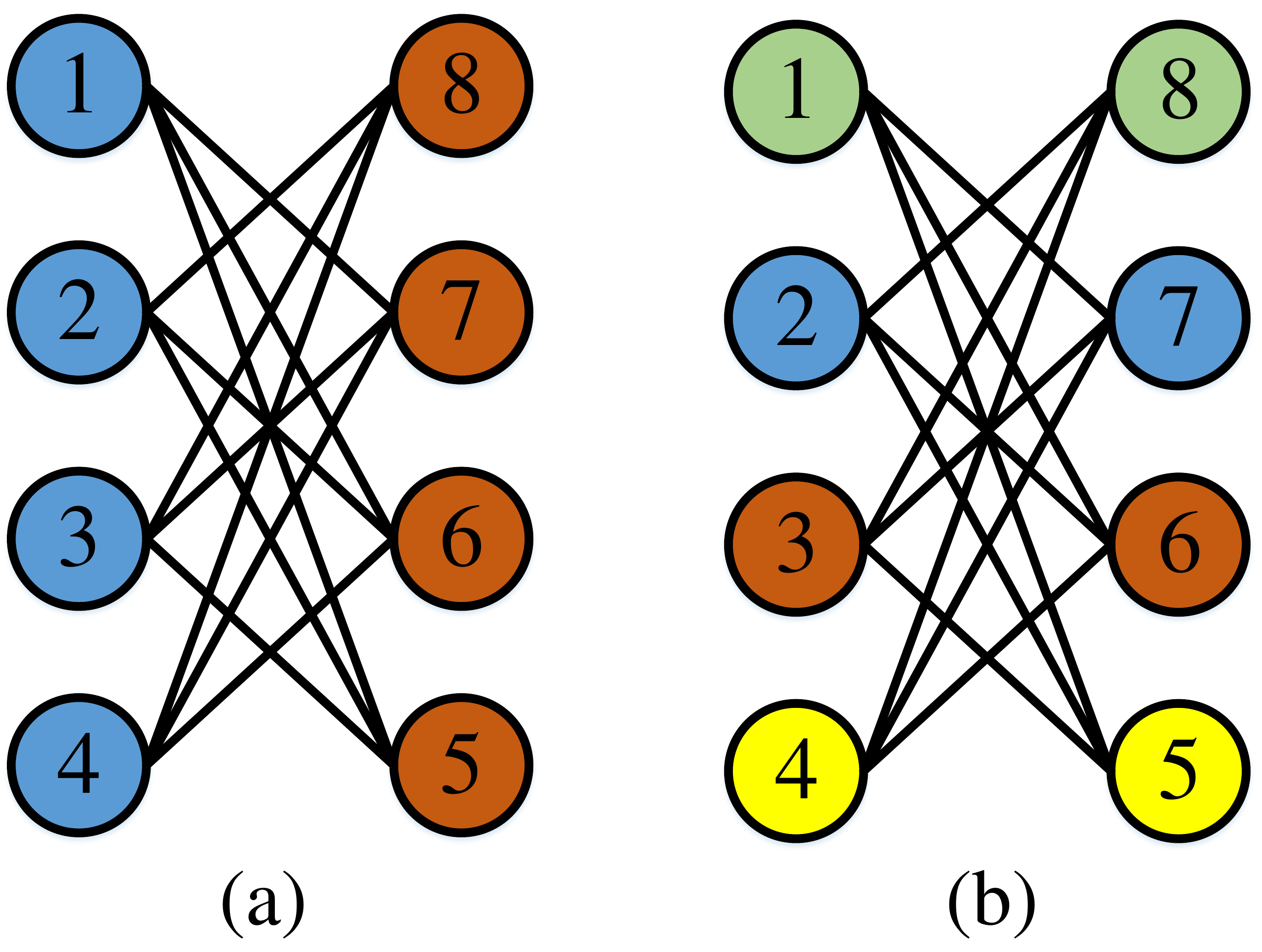}\\
  \caption{A greedy algorithm for a graph coloring problem (a) Graph connectivity (b) A solution from the greedy algorithm}
  \label{Greedy}
  \end{center}
\vspace{-1.em}
\end{figure}

This technique can be a valid technique that colors vertices in a graph because the constraint for the graph coloring is evaluated for each step. However, it is difficult to find the number of colors required for coloring the graph. The answer to this question is not trivial because the number of colors can depend on the order that we select which color to which vertex first. This technique requires a sequential process so that the performance overhead can be increased exponentially as the size of the graph increases. Fig. \ref{Greedy} shows an example of a greedy algorithm. As can be seen, the colors are numbered and at each time a new vertex is colored by the lowest number by considering the adjacent vertex constraints.

\begin{figure}[b]
  \begin{center}
  \includegraphics[width=3.5in]{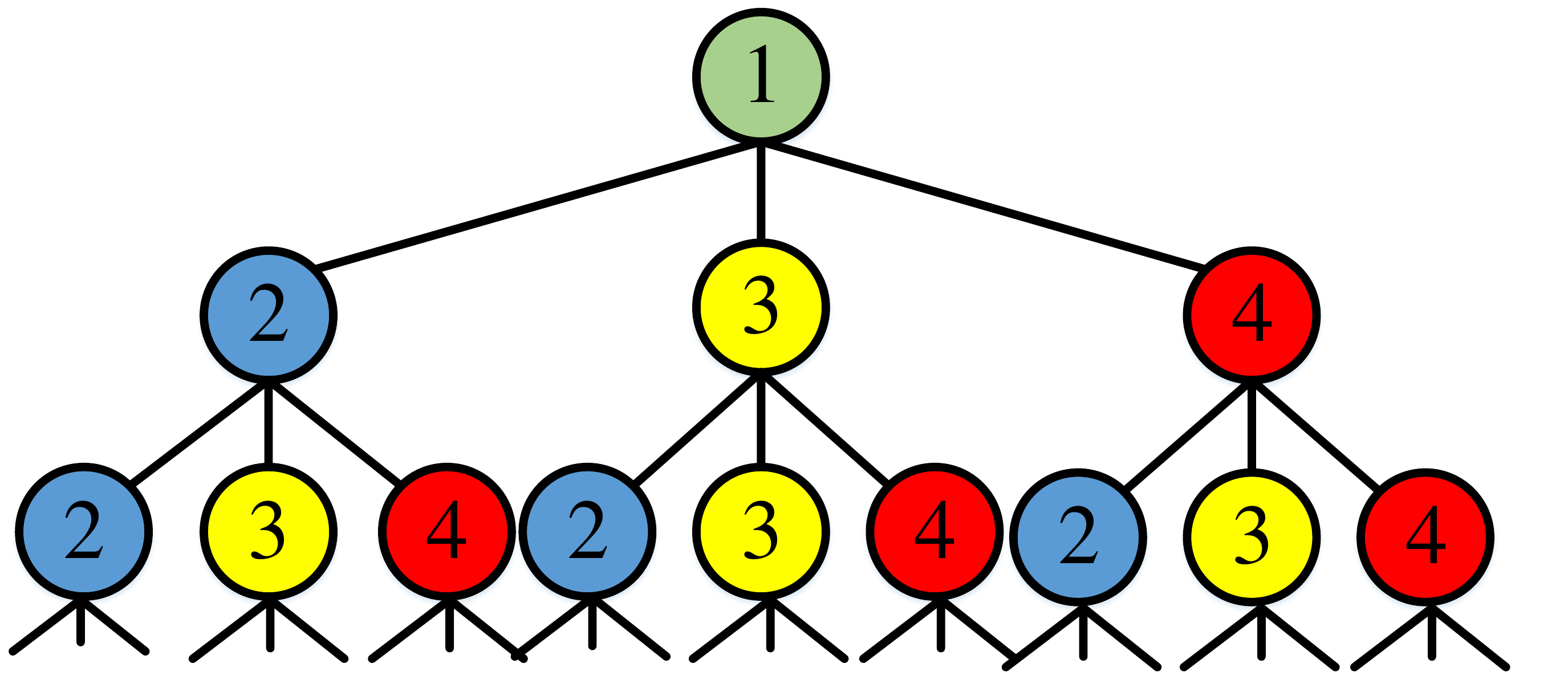}\\
  \caption{A backtracking algorithm for a graph coloring problem
}
  \label{Backtracking}
  \end{center}
\vspace{-1.2em}
\end{figure}

\subsubsection{Backtracking $k$-coloring algorithm}
This technique first forms a tree based on a graph problem and selects or assigns one color out of $k$ colors to the initial vertex (e.g., the root node in a tree)~\cite{ref24}. The algorithm then moves to the next node in the tree and use $k-1$ colors to assign the colors based on the constraints. This process is repeated until it reaches the last vertex in the tree. This process generates an initial sequence of colors.   The algorithm then moves starting from the last vertex to the top vertex to obtain another sequence based on the constraints. These processes will be repeated until the tree was completed for the graph. Then the best solution can be obtained by searching into the tree to find the best sequence of the colors with the minimum number of colors. Fig. \ref{Backtracking} shows an example of $k$-coloring based on Backtracking technique. 

\subsection{VQE algorithm for graph coloring problems}
In this section, we explain the VQE technique to solve multi-coloring graph problems. We also present a method that describes how to convert an optimization problem into a physical Hamiltonian matrix. 

For the graph $G(V,E)$ where $V(G) = \{1,2,...,n\}$ and $E(G) = \{e_{ij}\}$, let $x_{ij}=1$ if the color $j$ has been assigned to vertex $i$. From this constraint, the $k$-coloring problem can be modeled as,
\begin{equation}\label{k-colring_con1}
 \sum_{i,j=1} x_{ij} = 1, \;\;\;\; i =1,2,...,n ;\;\; j=1,2,...,k.
\end{equation}

\noindent Eq. (\ref{k-colring_con1}) shows that the color $j$ is assigned to node $i$. Next apply the second constraint which covers the color for the adjacent nodes. This constraint has been modeled as Eq. (\ref{k-colring_con2}),
\begin{equation}\label{k-colring_con2}
 x_{ip} + x_{jp} \leq {1}, \;\;\;\; p =1,2,...,k.
\end{equation}

\noindent Applying these two constraints to the cost function, the $k$-coloring problem can be formulated as a QUBO optimization problem. After adding the quadratic penalties, the cost function will be $P\sum_{i} (\sum_{j=1}^{n} a_{ij}x_{ij}-b_i)^2$, where $P$ is a large positive number. Bases on these constraints, the QUBO optimization problem can then be formulated as Eq. (\ref{QUBO_K_coloring}),
\begin{equation}\label{QUBO_K_coloring}
 min\;\; X^T Q X  + g^T X = \sum_{ijk} (P ( x_{ij} - 1 )^2 + P x_{ik}),
\end{equation}
where $x_{ij}$ is the node $i$ that the color $j$ has been assigned to it and $x_{ik}$ is referred to the nodes that are connected each other but they should not have the same color. The new set of the variables will be,
\begin{equation}\label{QUBO_K_coloring_variables}
 (x_1,x_2,x_3,..., x_{n\times k}) = (x_{11},x_{12},...,x_{1k}, x_{21}, ..., x_{n1},...,x_{nk}),
\end{equation}
where $n$ is the number of nodes and $k$ is the number of colors. As can be seen, the number of qubits required to solve a multi-coloring problem is $n\times k$. The derive $Q$ matrix is a hermitian matrix that can be converted to the required Hamiltonian matrix using Eq. (\ref{QUBO_sub}).

\section{ Results }\label{section-5}
In this section, we utilize both VQE and QAOA algorithms to demonstrate that our approach can find optimal solutions to three case studies for combinatorial problems: flight gate allocation, frequency allocation and register allocation. Our method has been tested and analyzed for the cases using IBM Qiskit platform~\cite{ref38}. In order to apply these algorithms, the combinatorial problems should be converted into the corresponding graphs and thereby the Hamiltonian matrices are constructed by applying Eq. (\ref{QUBO_sub}) with considering the constraints of node connectives in the graph.

\subsection{Case study 1: Flight gate allocation}
Demands for air travel have been growing with the improvement of the economy. As a result, airline management systems for assigning gate allocation for flights can be critical to meet the growing demand for air travel. For example, constructing more gates is a simple solution for solving gate allocation problems but it requires high costs and time. Hence, it is important to develop an efficient algorithm to increase the utilization of limited gates by optimizing gate allocation problems. Traditionally, different classical algorithms have been developed to solve small allocation problems. However, by increasing the number of flights and gates, solving the problems will take exponentially in time for classical computers~\cite{ref25,ref26}. By utilizing the advantage of quantum computers, it is feasible to develop an efficient algorithm for NISQ devices that can be applied to airline management gate allocation systems in the near future to optimize combinations of gates and flights in a reasonable amount of time.

\begin{figure}[t]
  \begin{center}
  \includegraphics[width=3.5in]{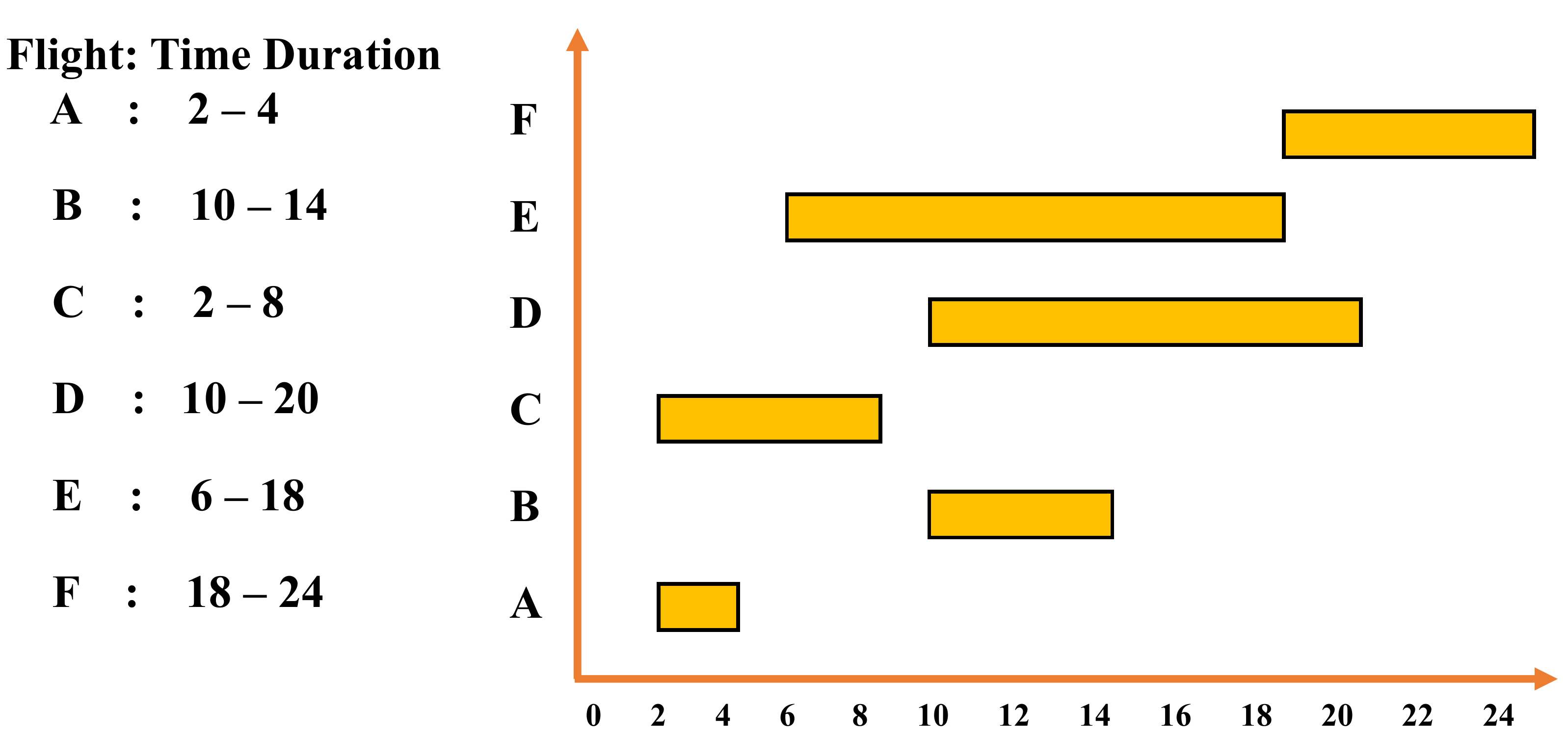}\\
  \caption{An example of flight gate allocation live range and live intervals}
  \label{flight}
  \end{center}
\vspace{-1em}
\end{figure}

\begin{figure}[t]
  \begin{center}
  \includegraphics[width=3in,height=1.4in]{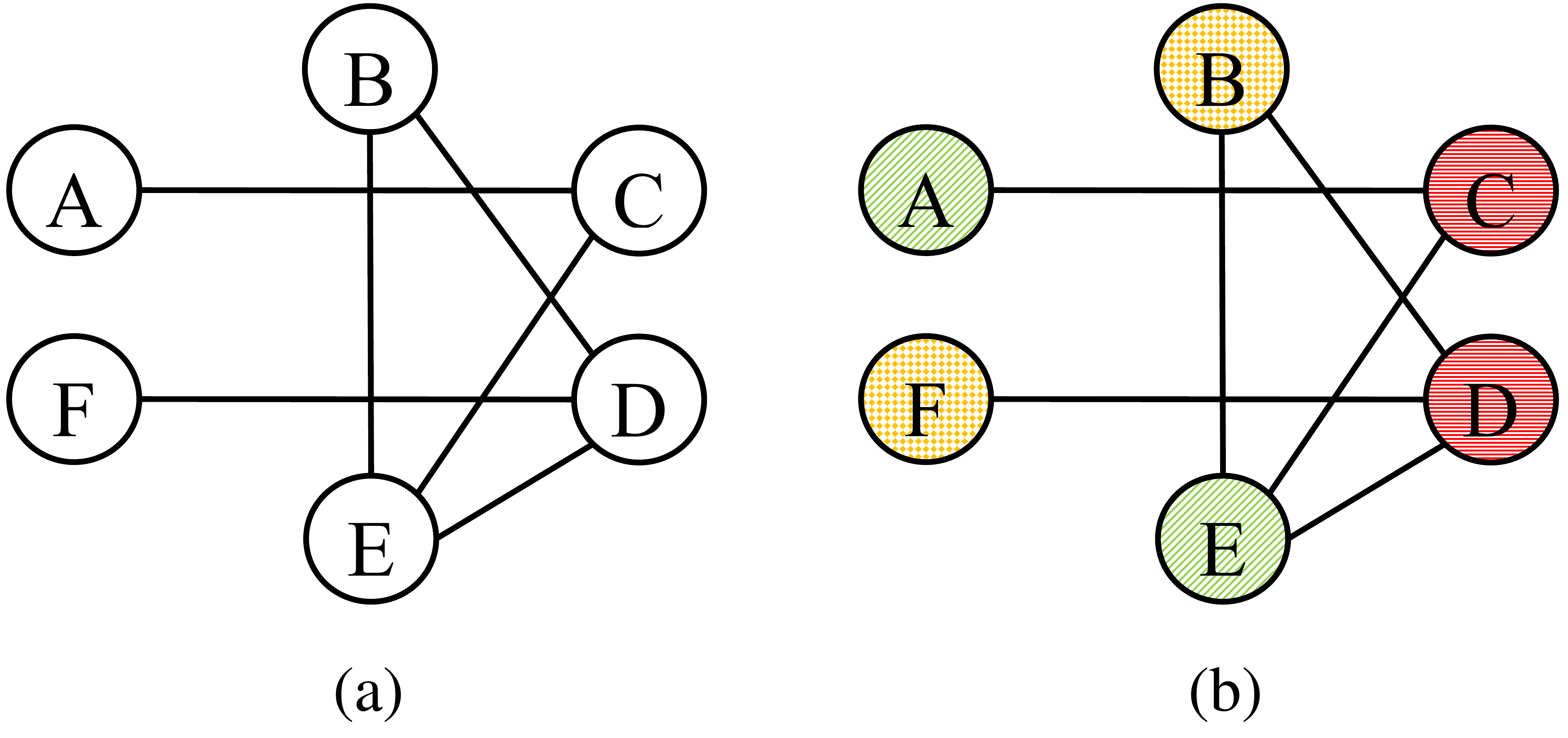}\\
  \caption{The flight gate allocation graph with 6 nodes and 3 colors (a) Graph connectivity (b) The solution from VQE algorithm}
  \label{flight_graph}
  \end{center}
\vspace{-1em}
\end{figure}

\begin{table}[b]
\caption{The combinatorial state solution $X_G$ from VQE algorithm for flight gate allocation graph. O: Orange, R: Red, G: Green}
\label{Flight_graph_table}

\setlength{\tabcolsep}{1.5\tabcolsep}
\centering
\begin{tabular}{|c|c|c|c|c|c|c|}
\hline
Graph's Nodes          & A           & B           & C           & D            & E           & F           \\ \hline
Colors                 & O    R    G & O    R    G & O    R    G & O    R    G  & O    R    G & O    R    G \\ \hline
$X_{G_{1\times 18}}$ & 0    0    1 & 1    0    0 & 0    1    0 & 0    1     0 & 0    0    1 & 1    0    0 \\ \hline
\end{tabular}
\end{table}

In this section, we present a method that first converts an example of gate assignment problems into a multi-coloring graph problem.  The main objective of our approach is to demonstrate that the VQE algorithm can be able to find an optimal solution for the small size of the gate allocation problem. Problems with a large number of gates and more time conflicts can be solvable using the same approach but it requires more computing resources and longer execution time to solve.

We consider 6 gates and 3 colors (i.e., the maximum number of time conflicts). Then, we convert the graph into a Hamiltonian matrix which is tested by using the VQE algorithm in the IBM Qiskit platform. Fig. \ref{flight} and Fig. \ref{flight_graph} (a) show the gates, the time intervals and the corresponding graph for the gate allocation for flights. As can be seen, the nodes are corresponding to the gates and the edges will be linked if the gates have time overlap interval among them. 

In our flight gate allocation case, the $Q$ and $g$ matrices with $P=4$ for the graph based on Eq. (\ref{QUBO_K_coloring}) can be calculated as, 

\begin{equation}\label{Qmatrix}
\setlength{\arraycolsep}{2.5pt} 
\medmuskip = 1mu 
\underset{18\times 18}{Q}=\begin{pmatrix*}[r]
   0& -4& -4&  0&  0&  0& -2&  0&  0&  0&  0&  0&  0&  0&  0&  0&  0&  0\\
  -4&  0& -4&  0&  0&  0&  0& -2&  0&  0&  0&  0&  0&  0&  0&  0&  0&  0\\
  -4& -4&  0&  0&  0&  0&  0&  0& -2&  0&  0&  0&  0&  0&  0&  0&  0&  0\\
   0&  0&  0&  0& -4& -4&  0&  0&  0& -2&  0&  0& -2&  0&  0&  0&  0&  0\\
   0&  0&  0& -4&  0& -4&  0&  0&  0&  0& -2&  0&  0& -2&  0&  0&  0&  0\\
   0&  0&  0& -4& -4&  0&  0&  0&  0&  0&  0& -2&  0&  0& -2&  0&  0&  0\\
  -2&  0&  0&  0&  0&  0&  0& -4& -4&  0&  0&  0& -2&  0&  0&  0&  0&  0\\
   0& -2&  0&  0&  0&  0& -4&  0& -4&  0&  0&  0&  0& -2&  0&  0&  0&  0\\
   0&  0& -2&  0&  0&  0& -4& -4&  0&  0&  0&  0&  0&  0& -2&  0&  0&  0\\
   0&  0&  0& -2&  0&  0&  0&  0&  0&  0& -4& -4& -2&  0&  0& -2&  0&  0\\
   0&  0&  0&  0& -2&  0&  0&  0&  0& -4&  0& -4&  0& -2&  0&  0& -2&  0\\
   0&  0&  0&  0&  0& -2&  0&  0&  0& -4& -4&  0&  0&  0& -2&  0&  0& -2\\
   0&  0&  0& -2&  0&  0& -2&  0&  0& -2&  0&  0&  0& -4& -4&  0&  0&  0\\
   0&  0&  0&  0& -2&  0&  0& -2&  0&  0& -2&  0& -4&  0& -4&  0&  0&  0\\
   0&  0&  0&  0&  0& -2&  0&  0& -2&  0&  0& -2& -4& -4&  0&  0&  0&  0\\
   0&  0&  0&  0&  0&  0&  0&  0&  0& -2&  0&  0&  0&  0&  0&  0& -4& -4\\
   0&  0&  0&  0&  0&  0&  0&  0&  0&  0& -2&  0&  0&  0&  0& -4&  0& -4\\
   0&  0&  0&  0&  0&  0&  0&  0&  0&  0&  0& -2&  0&  0&  0& -4& -4&  0
\end{pmatrix*}
\end{equation}

\begin{equation}\label{gmatrix}
\underset{1\times 18}{g^T} =\begin{pmatrix}
   4&  4&  4&  4&  4&  4&  4&  4&  4&  4&  4&  4&  4&  4&  4& 4&  4 
\end{pmatrix}
\end{equation}

\begin{figure}[b]
  \begin{center}
  \includegraphics[width=3.5in]{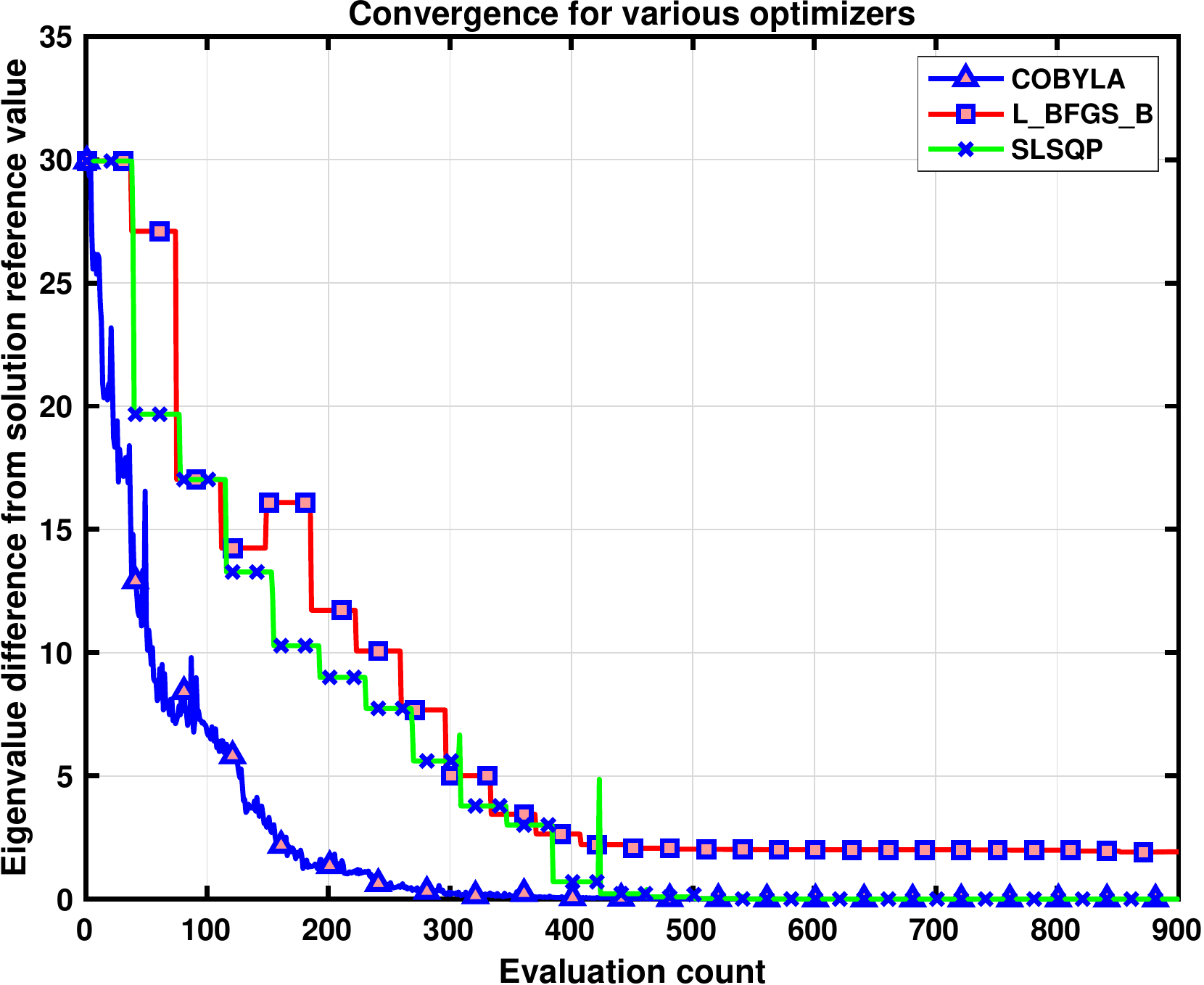}  \caption{Convergence of VQE algorithm with three different classical optimizers for flight gate allocation}
  \label{Comparison_flight}
  \end{center}
\vspace{-1em}
\end{figure}

We investigate different classical optimizers, considering three well-known methods such as COBYLA, L\_BFGS\_B, and SLSQP to optimize the parameters of the quantum circuits. Our simulation results demonstrate that all optimizers can find an optimal solution for the gate allocation problem. The assigned gates with different colors based on the results are describes in Table. \ref{Flight_graph_table} and shown is Fig. \ref{flight_graph} (b). As can be seen, the adjacent nodes have different colors so that our approach can optimize the gate allocation problem properly. Fig. \ref{Comparison_flight} shows the convergence of the classical optimizers as the optimization iteration increases. As can be seen the eigenvalue difference of $COBYLA$ optimizer has higher convergence speed.

\begin{figure}[t]
  \begin{center}
  \includegraphics[width=2.5in,height=2in]{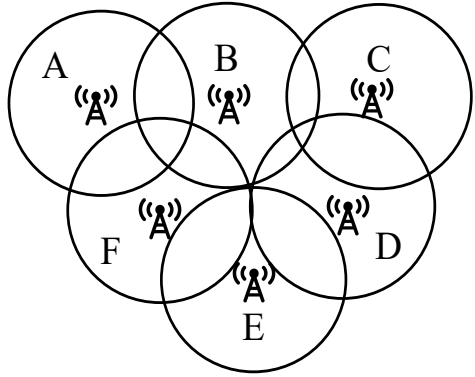}\\
  \caption{An interference model of the base stations}
  \label{bsestation}
  \end{center}
\vspace{-1em}
\end{figure}

\begin{figure}[t]
  \begin{center}
  \includegraphics[width=3in,height=1.4in]{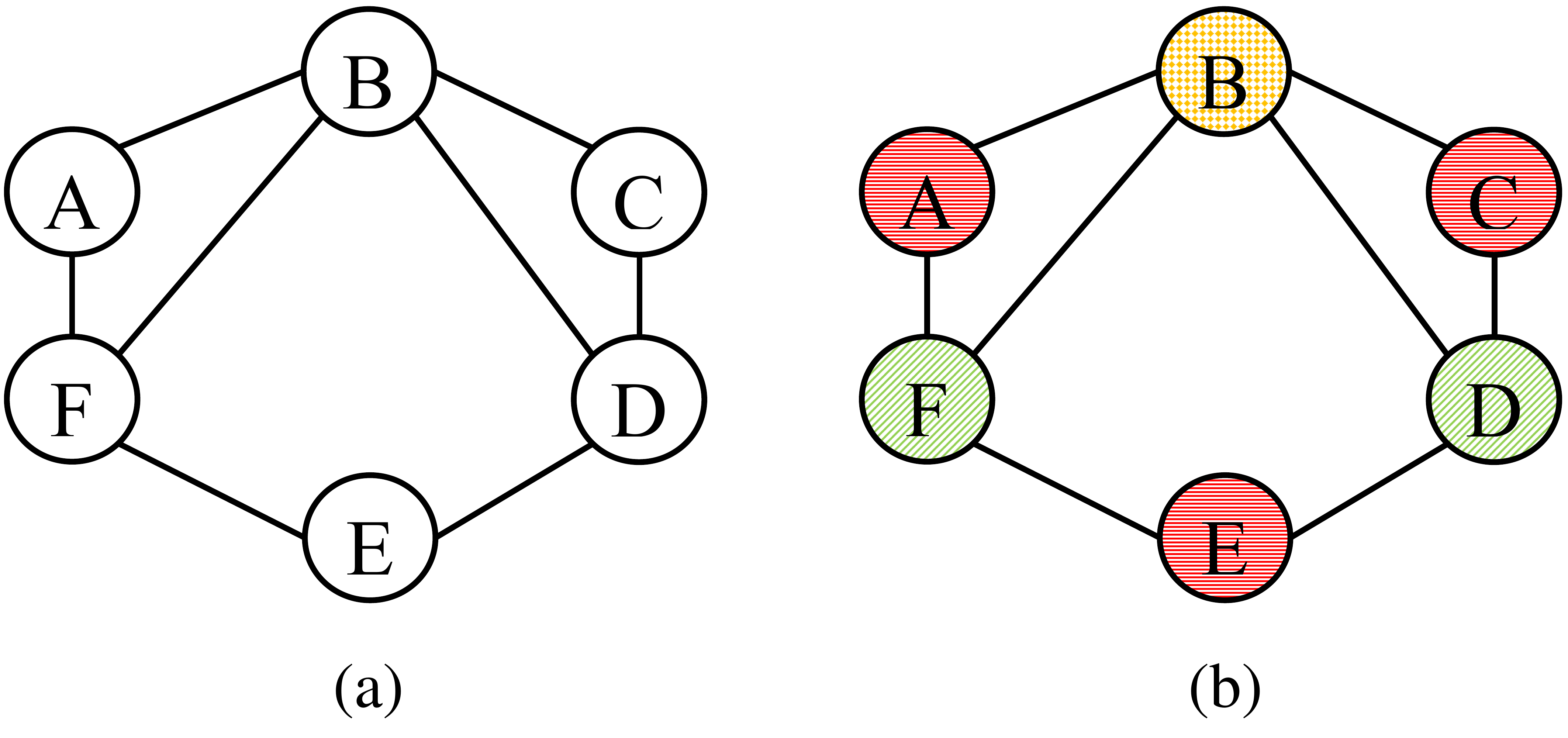}\\
  \caption{Frequency allocation graph with 6 nodes and 3 colors (a) Graph connectivity (b) A solution from VQE algorithm}
  \label{graph_frequency}
  \end{center}
\vspace{-1em}
\end{figure}

\subsection{Case study 2: Frequency allocation}
In this section, we present a method that utilizes the VQE technique to find an optimum solution to a Frequency assignment problem.  Improvement in wireless network technologies have contributed to the rapid growth of mobile networking and wireless communication applications for an expanding number of users.  This increase in the number of mobile users requires that the fixed size of the wireless frequency spectrum be shared among more and more users, causing issues of constraints in communications. This requires an efficient reuse of the frequencies and optimized allocations that assign frequencies or channels to base stations~\cite{ref27}. Such transmitter and receiver frequency assignment problems can result in 
communication overlaps and interference. Allocation algorithm need to be designed that can address networking constraints and avoid intermittent communication~\cite{ref28}.

\begin{figure}[t]
  \begin{center}
  \includegraphics[width=3.5in]{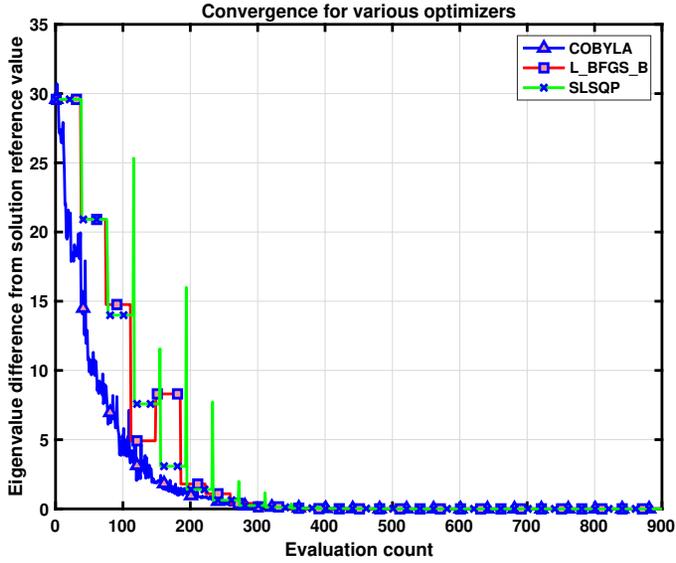}\\
  \caption{Convergence of VQE algorithm with three different classical optimizers for frequency allocation}
  \label{code_Optimizer}
  \end{center}
\vspace{-1em}
\end{figure}

\begin{table}[b]
\caption{The combinatorial state solution $X_F$ from VQE algorithm for frequency allocation graph. O: Orange, R: Red, G: Green}
\label{Frequency_graph_table}

\setlength{\tabcolsep}{1.5\tabcolsep}
\centering
\begin{tabular}{|c|c|c|c|c|c|c|}
\hline
Graph's Nodes        & A           & B           & C           & D           & E           & F           \\ \hline
Colors               & O    R    G & O    R    G & O    R    G & O    R    G & O    R    G & O    R    G \\ \hline
$X_{F_{1\times 18}}$ & 0    1    0 & 1    0    0 & 0    1    0 & 0    0    1 & 0    1    0 & 0    0    1 \\ \hline
\end{tabular}
\end{table}

As explained in the previous section, the classical approaches such as greedy technique, heuristic techniques, simulated annealing, and Tabu search algorithms have been proposed to find optimal solutions for the frequency assignment problems. However, rapid increase of the number of mobile users and the stations can make the problem  become an NP-hard problem because it is infeasible to solve it in polynomial time. 

Fig.~\ref{bsestation} shows an example of a frequency assignment problem with 6 base stations. In this case, we need to assign three different frequencies to the base stations in such a way that the adjacent base stations should not have the same frequency. The represented graph for this problem has been shown in Fig. \ref{graph_frequency} (a). The $Q$ and $g$ matrices can be calculated by Eq. (\ref{QUBO_K_coloring}). The problem has been tested and analyzed by using the VQE algorithm running on the IBM Qiskit platform. Different classical optimizers are used to optimize the quantum circuits and their results are presented in Fig. \ref{code_Optimizer}. As can be seen, all the optimizers have approximately the same convergence speed. The optimal solution for this problem has been shown in Tables. \ref{Frequency_graph_table}. The colored graph based on the results is shown in Fig. \ref{graph_frequency} (b).

\subsection{Case study 3: Register allocation}
Register allocation is one of the main problems in compiler theory. Many research efforts have been directed toward  developing different techniques to find an optimal solution for the problem. As more applications start running, the  resource management becomes more difficult to handle all the processes properly on time. Thus, this register allocation problem can become an NP-hard problem when a compiler needs to find an optimal solution to assign a large number of variables to the limited registers at the same time. Solving a register allocation problem requires vast amount of time and resources using classical computers~\cite{ref29}.

\begin{figure}[t]
  \begin{center}
  \includegraphics[width=3.5in]{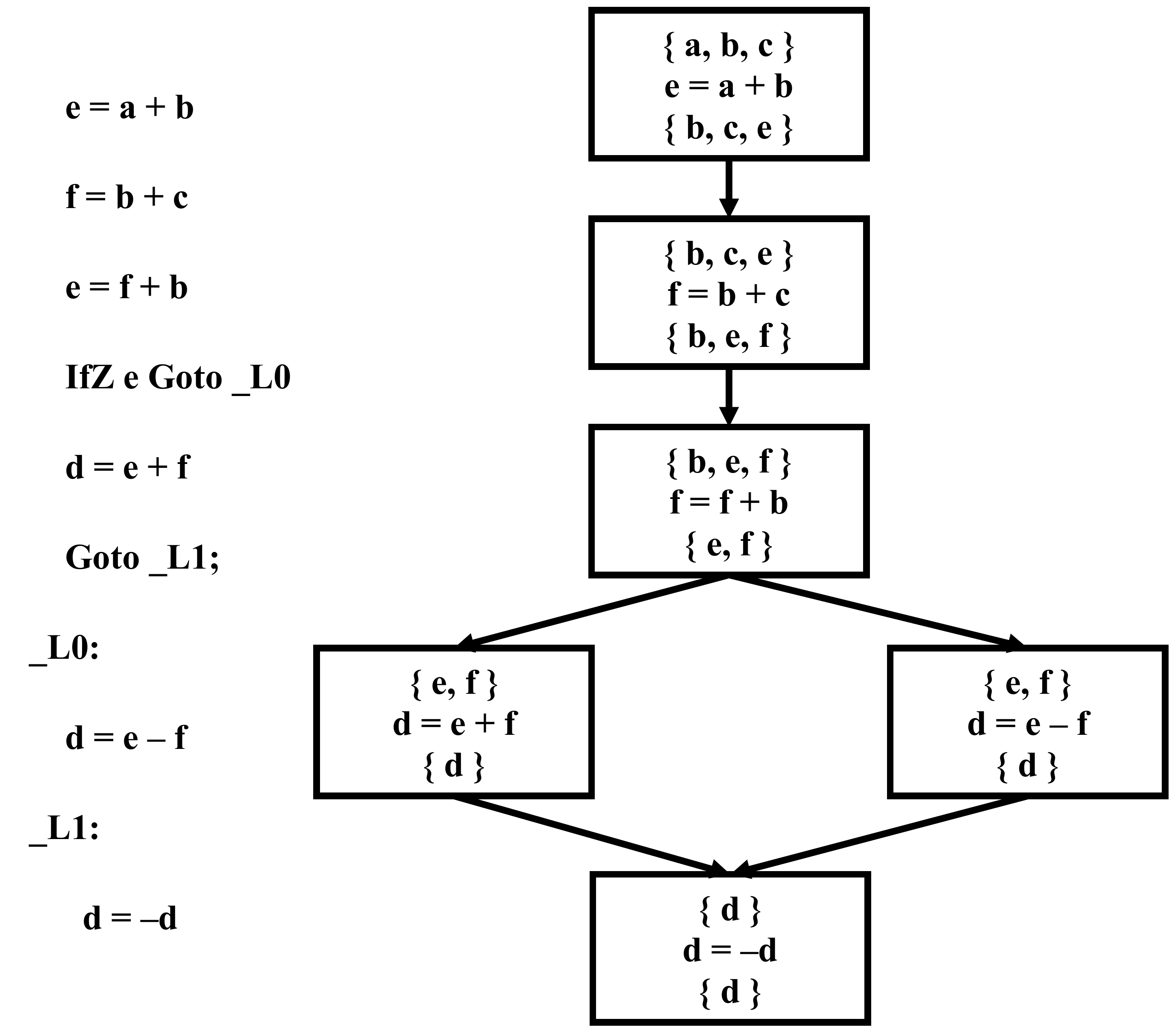}\\
  \caption{An example of register allocation live range and live intervals}
  \label{reg_chart}
  \end{center}
\vspace{-1em}
\end{figure}

In register allocation, the main task is to optimize how a compiler can efficiently assign a large number of variables to a limited number of CPU registers in order to maximize the CPU utilization and minimize application run times. Recent compilers can solve register allocation and instruction scheduling problems separately using heuristic optimization algorithms. In an instruction scheduling algorithm, a code generator creates intermediate code which can be used to reduce the total latency~\cite{ref30}. However, the final assembly translation can be more complicated because the scheduling instructions need to be reordered in a separate way to enhance the total latency. 

\begin{table}[b]
\caption{The combinatorial state solution $X_R$ from VQE algorithm for register allocation graph. O: Orange, R: Red, G: Green}
\label{Register_graph_table}

\setlength{\tabcolsep}{1.5\tabcolsep}
\centering
\begin{tabular}{|c|c|c|c|c|c|c|}
\hline
Graph's Nodes        & A           & B           & C           & D           & E           & F           \\ \hline
Colors               & O    R    G & O    R    G & O    R    G & O    R    G & O    R    G & O    R    G \\ \hline
$X_{R_{1\times 18}}$ & 0    0    1 & 1    0    0 & 0    1    0 & 1    0    0 & 0    1    0 & 0    0    1 \\ \hline
\end{tabular}
\end{table}

\begin{figure}[t]
  \begin{center}
  \includegraphics[width=3in,height=1.4in]{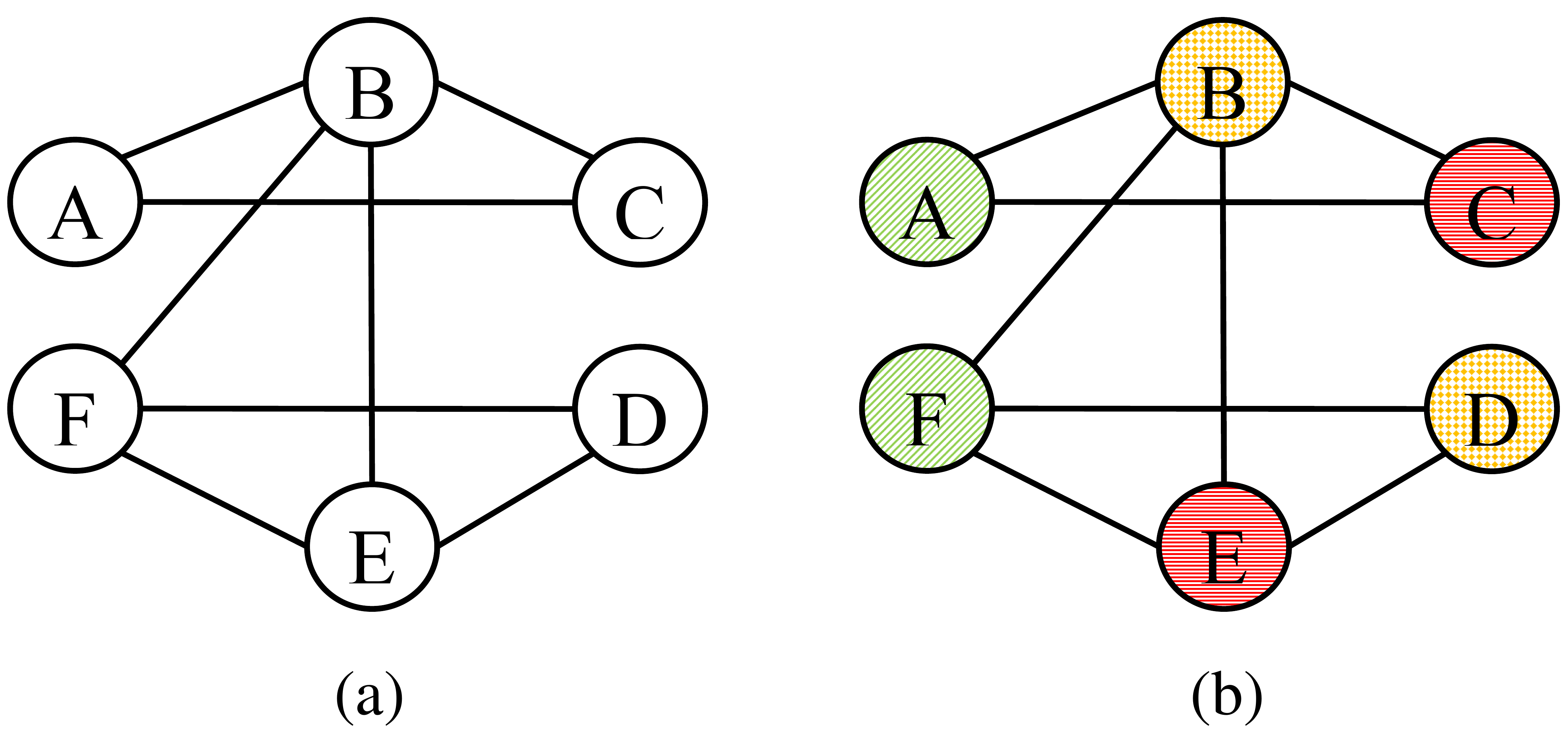}\\
  \caption{Register allocation graph with 6 nodes and 3 colors (a) Graph connectivity (b) A solution from VQE algorithm}
  \label{Reg_graph}
  \end{center}
\vspace{-1em}
\end{figure}

The goal of register allocation is to regenerate the code before the program execution time in such a way that it contains fewer number of intermediate variables than the number of available registers.  This compilation process will not change the behaviors of the code but it can improve the utilization of registers so that an optimal register allocation algorithm can reduce the execution time of the application by avoid unnecessary register accesses. Swapping in and out values between register and memory require more CPU clock cycles because accessing the memory needs to use most of the hardware parts such as memory buses, cache controllers which consumes more running CPU cycles. 

A three-address code (TAC) example for register allocation problem is described in Fig. \ref{reg_chart}. The problem can be converted to a graph problem by considering the total intermediate values as the nodes of the graph. Two nodes will be connected if the values of both temporaries are needed to use at the same time. If there is no edge among the temporaries, then different physical registers can be assigned to them. Fig. \ref{Reg_graph} (a) shows the corresponding graph for the problem.

\begin{figure}[t]
  \begin{center}
  \includegraphics[width=3.5in]{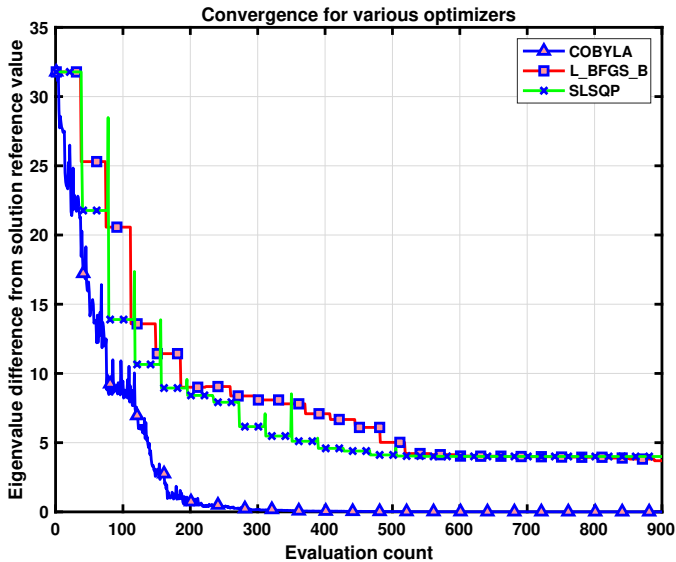}\\
  \caption{Convergence of VQE algorithm with three different classical optimizers for register allocation}
  \label{Reg_optimizers}
  \end{center}
\vspace{-1em}
\end{figure}

We simulate our approach on the IBM Qiskit platform that utilizes the VQE algorithm to solve the register allocation problem. The results of VQE algorithm with different classical optimizers are shown in Fig. \ref{Reg_optimizers}. All the optimizers are converged to the optimal solution for the problem with different convergence speed. Among them $COBYLA$ optimizer has better convergence speed. The optimal solution for this problem is presented in Table. \ref{Register_graph_table} and the colored graph based on the solution is described in Fig. \ref{Reg_graph} (b). As can bee seen form the colored graph, the adjacent nodes are assigned with different colors.

\begin{figure}[t]
  \begin{center}
  \includegraphics[width=4.3in, height= 2in]{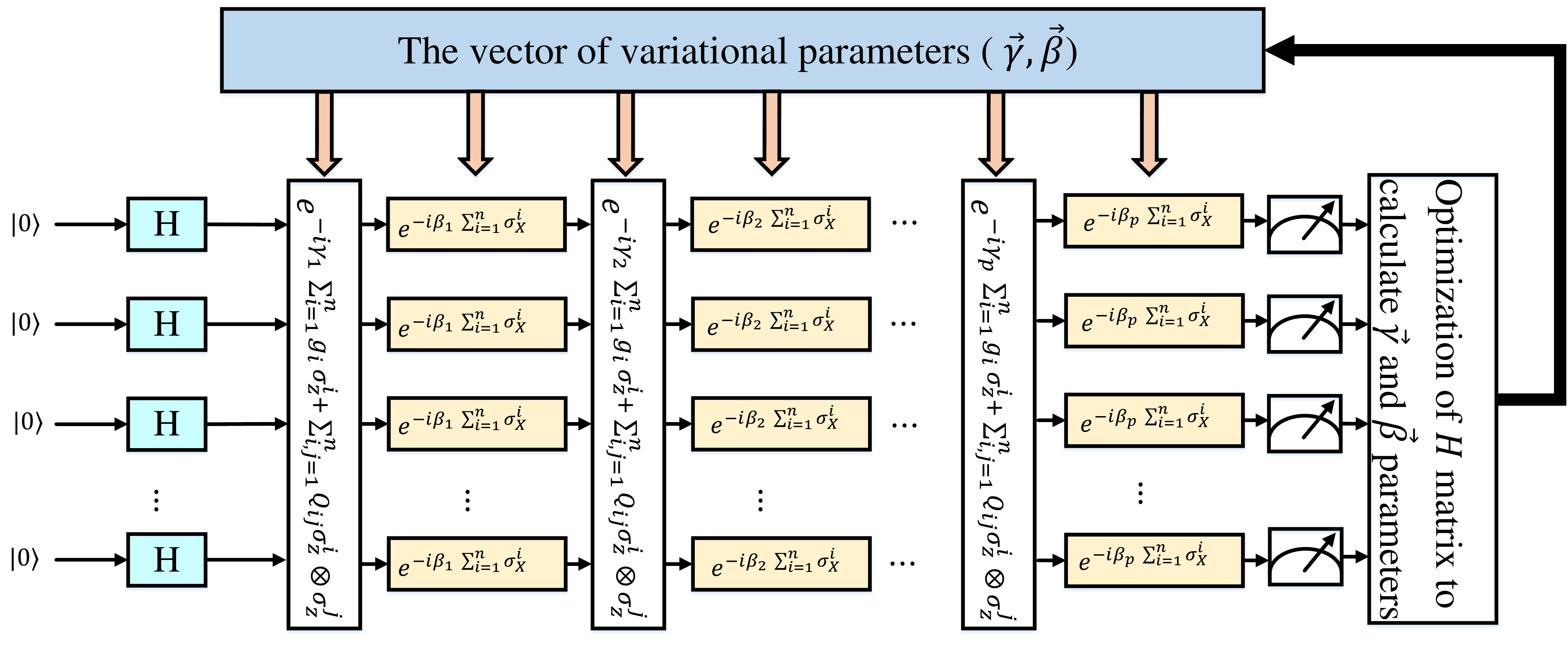}\\
  \caption{The quantum/classical circuit diagram for QAOA algorithm}
  \label{QAOA_Circuit}
  \end{center}
\vspace{-1em}
\end{figure}

\section{Discussion}\label{section-6}
In this section, we investigate the quantum approximate optimization algorithm (QAOA) to solve the combinatorial problems. QAOA is a hybrid quantum-classical approach that can be particularly useful for solving combinatorial optimization problems. In QAOA, the variational form is derived from the final Hamiltonian matrix $H$ that can be constructed by the problem's objective function. This technique uses an adiabatic evolution to find the ground state of the final Hamiltonian $H$ matrix. As the depth of the quantum circuits increases, the approximation ratio with respect to the ground state (an optimal solution) can be guaranteed. The circuit diagram of the QAOA algorithm is shown in Fig. \ref{QAOA_Circuit}. By considering the Eq. \ref{QUBO}, the variational quantum circuit can be constructed as,
\begin{equation}\label{QAOA_Ugama}
 U_c(\gamma) = e^{-i\gamma \big(\sum_{i=1}^{n} g_i \sigma_z^i + \sum_{i,j=1}^{n} Q_{ij} \sigma_z^i \bigotimes \sigma_z^j\big) }
\end{equation}
\begin{equation}\label{QAOA_Ubeta}
 U_B(\beta) = e^{-i\beta \sum_{i=1}^{n} \sigma_X^i},
\end{equation}

\begin{figure}[b]
  \begin{center}
  \includegraphics[width=3.5in]{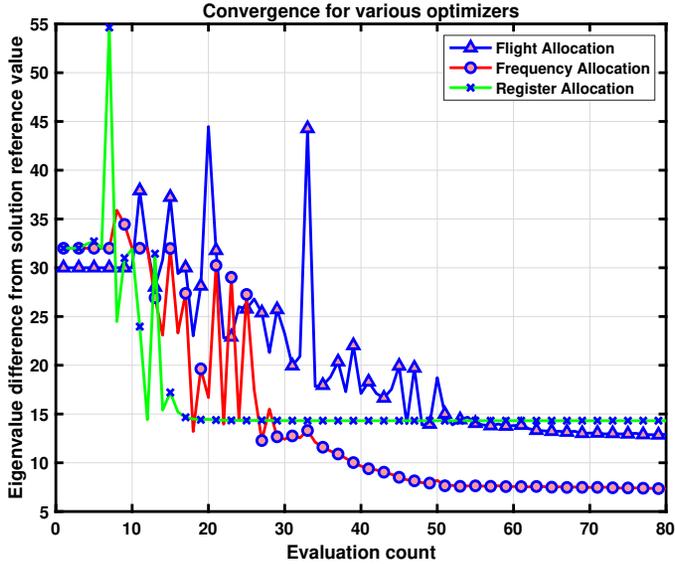}\\
  \caption{Convergence of QAOA algorithm with COBYLA optimizer for three case studies}
  \label{qaoa_cobyla}
  \end{center}
\vspace{-1em}
\end{figure}

\noindent where $\gamma$ and $\beta$ $\in\mathbb{R}^p$ are the variational parameters that should be optimized by the classical optimizer and $p$ is the considered as the depth of the circuit. The optimization process can be done by calculating the expected values through multiple observations. As can be seen, the QAOA algorithm uses $2p$ variational parameters to find the solution of the combinatorial problem. $U_c(\gamma)$ can be constructed by using two CNOT gates and one rotation $Z$ gate if $Q_{ij}\neq 0$. Also, $U_b(\beta)$ requires a single rotation $X$ gate. In general, the variational quantum circuit in QAOA with $n$ and $p$ requires $O(n^2p)$ single rotation gates and $O(n^2p)$ CNOT gates.
\begin{equation}\label{QAOA_Ubeta}
 U(\beta, \gamma) = \left(\prod_{i = 1}^{p} \;U_B(\beta)\;\; U_c(\gamma)\right) H^{\bigotimes n}.
\end{equation}

In this section, the QAOA algorithm was applied to the three combinatorial case studies discussed in section ~\ref{section-4}. The $Q$ and $g$ matrices for each case were generated and used to run the QAOA algorithm. The COBYLA optimizer was used to perform the optimization of parameters. The results of the problems are shown in Fig. \ref{qaoa_cobyla}. As can be seen, the QAOA algorithm with COBYLA optimizer was converged to the optimum solutions for each case study. Unlike the VQE technique, the QAOA algorithm guarantees the approximation solution for the combinatorial optimization problem. Furthermore, if the depth of the circuit $p$ goes to $\infty$, the QAOA algorithm will be converged to the exact solution of the combinatorial problem. The approximated solutions for different case studies derived by QAOA algorithm are presented in Table. \ref{QAOA_table}. As can be seen, from the results the QAOA algorithm is capable to solve the combinatorial problems and derive the optimum solutions.

\begin{table}[t]
\caption{The combinatorial state solution from QAOA algorithm for all case studies. O: Orange, R: Red, G: Green}
\label{QAOA_table}

\setlength{\tabcolsep}{1.5\tabcolsep}
\centering
\begin{tabular}{|c|c|c|c|c|c|c|}
\hline
Graph's Nodes & A           & B           & C           & D           & E           & F           \\ \hline
Colors        & O    R    G & O    R    G & O    R    G & O    R    G & O    R    G & O    R    G \\ \hline
$X_{G_{1\times 18}}$          & 0    0    1 & 1    0    0 & 1    0    0 & 0    0    1 & 0    1    0 & 0    1    0 \\ \hline
$X_{F_{1\times 18}}$          & 0    0    1 & 0    1    0 & 1    0    0 & 0    0    1 & 0    1    0 & 1    0    0 \\ \hline
$X_{R_{1\times 18}}$          & 0    1    0 & 1    0    0 & 0    0    1 & 1    0    0 & 0    0    1 & 0    1    0 \\ \hline
\end{tabular}
\end{table}

\section{Conclusion}\label{section-7}
This paper presents an approach to solve multi-coloring graph problems using the hybrid quantum algorithms that can be able to take the advantage of both quantum computers and classical computing resources. Coloring a graph using the minimum number of colors is critical in the field of combinatorial optimization researches because it is an $N$-$P$ hard problem but their solutions can be applied to the variety of real applications. Although three different classical approaches such as exact solution algorithm, approximation  algorithms and heuristic algorithm have been used to solve the combinatorial optimization problem, it still requires exponential processing time as the number of nodes and edges in a graph increase rapidly. To address the problem, we utilize the variational quantum eigensolver (VQE) and quantum approximate optimization algorithm (QAOA) algorithms not only to find the optimal solutions for the combinatorial problems but also reduce the process time. This paper presents an approach to derive the objective function from the graph coloring problem with constraints and then converts the function into the appropriate Ising model. We apply our approach to find the optimal solutions for three different practical applications. Our simulation results demonstrate that our method with VQE and QAOA algorithms can be able to find optimal solutions of the combinatorial optimization problems by utilizing the quantum properties and classical optimizing techniques.


\bibliographystyle{unsrtnat}
\bibliography{Bibliography}

\end{document}